\documentclass[a4paper]{article}

\usepackage{INTERSPEECH2019}
\usepackage{amsmath,graphicx,amssymb}
\usepackage[export]{adjustbox}
\usepackage{subfig}
\usepackage{color}
\usepackage{enumitem}
\usepackage{multirow}
\PassOptionsToPackage{hyphens}{url}\usepackage{hyperref}

\hypersetup{colorlinks=true}

\title{Multimodal Target Speech Separation with Voice and Face References}
\name{Leyuan Qu, Cornelius Weber, Stefan Wermter}
\address{
  Knowledge Technology, Department of Informatics, University of Hamburg\\
  Vogt-Koelln-Str. 30, 22527 Hamburg, Germany}
\email{\{qu,weber,wermter\}@informatik.uni-hamburg.de}
\begin{document}

\maketitle

\begin{abstract}

\noindent Target speech separation refers to isolating target speech from a multi-speaker mixture signal by conditioning on auxiliary information about the target speaker. Different from the mainstream audio-visual approaches which usually require simultaneous visual streams as additional input, e.g. the corresponding lip movement sequences, in our approach we propose the novel use of a single face profile of the target speaker to separate expected clean speech. We exploit the fact that the image of a face contains information about the person's speech sound. Compared to using a simultaneous visual sequence, a face image is easier to obtain by pre-enrollment or on websites, which enables the system to generalize to devices without cameras. To this end, we incorporate face embeddings extracted from a pretrained model for face recognition into the speech separation, which guide the system in predicting a target speaker mask in the time-frequency domain. The experimental results show that a pre-enrolled face image is able to benefit separating expected speech signals. Additionally, face information is complementary to voice reference and we show that further improvement can be achieved when combing both face and voice embeddings\footnote{Web demo: https://leyuanqu.github.io/INTERSPEECH2020/}.

\end{abstract}

\section{Introduction}

\begin{figure}[th]
  \centering
  \includegraphics[width=\linewidth]{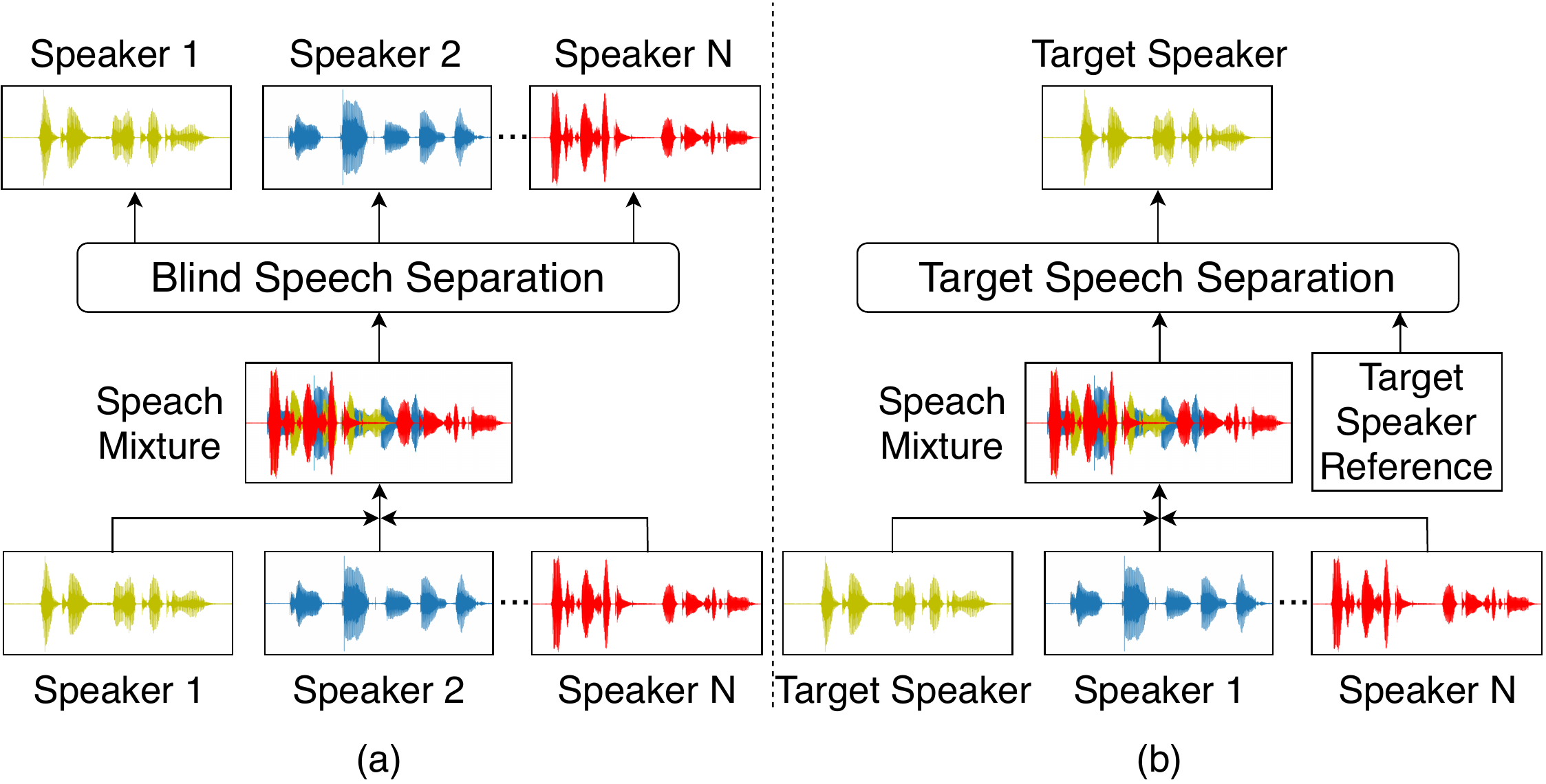}
  \caption{Comparison of (a) blind speech separation and (b) target speech separation.}
  \label{fig:target-ss}
\end{figure}

Speech separation aims to recover a clean speech signal from a mixture signal produced by multiple speakers simultaneously, e.g. in a cocktail party environment. Despite the significant progress on speech separation technologies over the past few years ~\cite{hershey2016deep, afouras2018conversation, Ephrat_2018}, the permutation problem is still challenging for the speech signal processing community. The permutation problem arises from label ambiguity --- the arbitrary order of multi-output --- which leads to an inconsistent gradient update and makes a neural network hard to converge during training. According to whether additional information can be available, approaches for solving the problem can be mainly divided into two categories: blind speech separation and target speech separation. The blind speech separation task is to isolate a clean output for each individual source signal without any other information about the observed speech mixture, as shown in Figure \ref{fig:target-ss} (a). To alleviate the permutation problem, deep clustering ~\cite{hershey2016deep} and its variant, a deep attractor network ~\cite{chen2017deep}, were proposed to disambiguate the label permutation. Permutation invariant training ~\cite{yu2017permutation} was presented to predict a best label permutation, whereas the unknown number of sources and invalid outputs are still big challenges in this direction ~\cite{luo2020separating}.

% assumes that 
Different from blind speech separation, target speech separation only recovers the desired single signal guided by auxiliary information, e.g. source directions or target speaker identity, as shown in Figure \ref{fig:target-ss} (b). By leveraging the target speaker reference, target speech separation avoids the permutation problem and is independent of the number of source speakers, since there is only one output per time in this case. 

More recently, multimodal audio-visual approaches have shown impressive results in target speech separation and attracted a lot of attention from the computer vision community, for instance, utilizing the lip movement sequences in videos to predict target time-frequency masks or directly generate the target waveform.

Inspired by VoiceFilter~\cite{wang2018voicefilter} which performed target speech separation with speaker voice embeddings and achieved good performance, in this paper, we extend the audio-only VoiceFilter to the audio-visual domain and explore to what extent the visual modality (face embedding) can benefit target speech separation, compared to only using voice as reference inputs. Additionally, previous audio-visual methods strictly require simultaneous visual streams and highly depend on the visual temporal information. This is hard to meet in most real-world cases, because the speaker’s mouth may be concealed by microphone ~\cite{afouras2019my} or be undetectable sometimes. Therefore, it is difficult to generalize to devices without cameras. To solve this problem, we propose to integrate the speaker face information into the system, which can be enrolled beforehand and easily applied to more challenging non-camera scenarios.

\begin{figure*}[t]
  \centering
  \includegraphics[width=\linewidth]{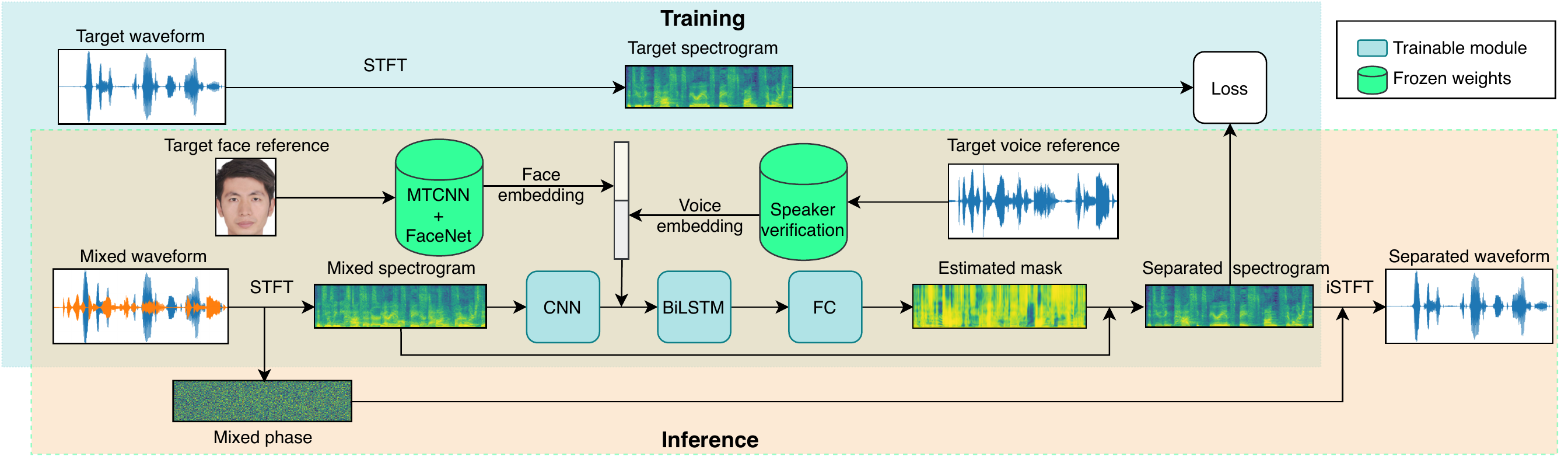}
  \caption{Overview of the proposed target speech separation architecture. The model receives inputs, i.e. the mixed spectrogram, the face embedding and/or the voice embedding to predict a target speaker time-frequency mask which is used to estimate the target spectrogram.}
  \label{fig:model_architecture}
\end{figure*}

\section{Related Work}

\noindent\textbf{Target speech separation:} Researchers working in this field try to inform models to only concentrate on the target output utilizing auxiliary information, such as source directions ~\cite{perotin2018multichannel}, spatial features ~\cite{chen2018multi}, speaker identity for multi-channel ~\cite{zmolikova2017speaker} and single-channel ~\cite{delcroix2018single} setups, speaker profile for both the target and competing speakers ~\cite{xiao2019single}, and so on.

Recently, there has been a growing interest in using multimodal audio-visual methods in target speech separation. Rather than only refining a target spectrogram and reconstructing a waveform with the phase from noisy speech, Afouras \textit{et al.} ~\cite{afouras2018conversation} use convolutional neural networks for both magnitude and phase estimation conditioning on lip regions in the corresponding video. Furthermore, considering the fact that the visual streams may be corrupted from realistic environments, for example, when the mouth region of the speaker is occluded by a microphone, Afouras \textit{et al.} ~\cite{afouras2019my} combine lip movement and self-enrolled voice representation to improve the robustness of the proposed system and to prevent visual modality domination. In a similar work, Ephrat \textit{et al.} ~\cite{Ephrat_2018} validate the effectiveness of using the whole face embedding, instead of just the lip area ~\cite{afouras2018conversation, afouras2019my}, to learn the target speaker magnitude mask based on a large-scale dataset in real-world scenarios. Different from previous works focusing on time-frequency masks, Wu \textit{et al.} ~\cite{wu2019time} directly estimate a raw waveform in the time domain by extending the audio-only (single-modal) TasNet~\cite{luo2018tasnet} into the audio-visual (multimodal) domain.  Gu \textit{et al.} ~\cite{gu2020multimodal} explore the effectiveness of using more information, i.e. speaker spatial location, voice characteristics, and lip movements, in target speech separation. A factorized attention mechanism was introduced to dynamically weigh the three kinds of additional information at the embedding level. Different from previous audio-visual works using corresponding video streams as auxiliary information, the objective of this paper is to investigate the benefit of the pre-enrolled face image for target speech separation.

\noindent\textbf{Learning associations between faces and voices:}
Inspired by the finding by neuroscientists~\cite{belin2004thinking, mavica2013matching} and psychologists~\cite{bruce1986understanding, schweinberger2007hearing} that there is a strong relationship between faces and voices and sometimes humans can even infer what one's voice sounds like by only seeing the face, or vice versa, researchers in computer science have conducted a large number of studies on learning face and voice association that can be mainly divided into two categories: crossmodal representation and joint/shared representation.

Work on the crossmodal representation has led to the possibility of generating one modality from another, e.g. reconstructing human faces by only conditioning on speech signals. Oh \textit{et al.} ~\cite{oh2019speech2face} design neural networks to directly map speech spectrogram to face embeddings which were pretrained for face recognition, then decoded the predicted face representation to canonical face images with a separate reconstruction model. Wen \textit{et al.} ~\cite{wen2019face} utilize generative adversarial networks (GAN) to generate human faces from the output of a pretrained voice embedding network. Instead of using a pretrained network, Choi \textit{et al.} ~\cite{choi2020inference} build speech and face encoders on a speech to face identity matching task, and train the encoders and a conditional generative adversarial network end to end to conduct face generation.

Researchers working on joint representation learning attempt to find a joint or sharing face-voice embedding space for tasks of crossmodal biometric retrieval or matching, e.g. searching a corresponding face image via a given speaker voice. Nagrani \textit{et al.} ~\cite{nagrani2018learnable} adopt a self-supervision training strategy to learn joint face and voice embeddings from videos without requiring any labelled data. Kim \textit{et al.} ~\cite{kim2018learning} introduce triplet loss to learn overlapping information between faces and voices by using VGG16 ~\cite{simonyan2014very} and SoundNet ~\cite{aytar2016soundnet} for visual and auditory modality respectively. Wen \textit{et al.} ~\cite{wen2018disjoint} propose DIMNet to leverage identity-sensitive factors, such as nationality and gender, as supervision signals to learn a shared representation for different modalities. Based on the strong association between faces and voices, we propose to utilize face embedding to guide models in tracking desirable auditory output.
\section{Model Architecture}
As shown in Figure \ref{fig:model_architecture}, our proposed model contains three neural networks: a pretrained FaceNet for face embedding extraction, a pretrained speaker verification net for voice embedding extraction, and a mask estimation net (the trainable modules) for target speaker mask prediction.

\subsection{Face embedding net}
The face embedding net is based on a Multi-task CNN (MTCNN) ~\cite{zhang2016joint} and FaceNet ~\cite{schroff2015facenet} used in a sequence. Before feeding the original face images into FaceNet, an MTCNN is used for face detection, since the MTCNN performs better in some hard conditions, such as partial occlusion and silhouettes. We crop only the face region and reshape all faces to 160x160 size for face embedding extraction. FaceNet directly learns a unified embedding for different tasks, for example face recognition and face verification, and achieves good results on different benchmarks. In this paper, we use FaceNet Inception-ResNet-v1 in Pytorch\footnote{https://github.com/timesler/facenet-pytorch}. The model is pretrained on the VGGFace2 dataset and achieves 99.65\% accuracy on the evaluation set. 

% by directly map face images to a Euclidean space

\subsection{Voice embedding net}
The voice embedding net is based on the model proposed by Wan \textit{et al.} ~\cite{wan2018generalized} for speaker verification, which consists of 3 LSTM layers with 768 nodes in each layer and one linear layer with 256-dimensional outputs. A generalized end-to-end loss was performed to cluster the utterances from the same class closer while increasing the distance between utterances from different classes during training. The pretrained model\footnote{https://github.com/mindslab-ai/voicefilter} used in our paper is trained on the VoxCeleb2 ~\cite{chung2018voxceleb2} dataset with thousands of speakers. The input spectrogram is extracted using the Short Time Fourier Transform (STFT) with a 40ms hop length and a 80ms window size. The model achieves 7.4\% equal error rate on the VoxCeleb1 test dataset (first 8 speakers).

\subsection{Mask estimation net}

The mask estimation network (the trainable modules in Figure \ref{fig:model_architecture}) is to predict a target speaker mask in the time-frequency domain, which is heavily inspired by VoiceFilter ~\cite{wang2018voicefilter} and the architecture proposed by Wilson \textit{et al.} ~\cite{wilson2018exploring}. As shown in Table \ref{tab:Mask_estimation_net}, the network begins with 7 Conv2D layers with different kernel sizes to capture the variations in time and frequency. Stacked dilated factors enable the network to have larger receptive fields. The output from the last CNN layer is concatenated with voice or/and face embeddings (repeated N times where N is the dimension of the spectrogram in time) as the input of the following bidirectional LSTMs layers. Two fully connected (FC) layers are used to map the high-dimensional outputs from LSTM to the dimension of spectrogram frequency. We use batch normalization and ReLU activation between each layer and a sigmoid function at the output layer. The separated spectrogram is obtained by multiplying the estimated mask and the mixed input. During inference, the separated waveform is reconstructed by the inverse STFT with the phase from the noisy mixture.

\begin{table}[th]
  \caption{Configuration of mask estimation network. Kernel is the kernel size in time and frequency. Dilation is the dilation factor in time and frequency.}
  \label{tab:Mask_estimation_net}
  \centering
    \begin{tabular}{cccccc}
    \toprule
    \multirow{2}{*}{\textbf{Layer}} & \multicolumn{2}{c}{\textbf{Kernel}} & \multicolumn{2}{c}{\textbf{Dilation}} & \multirow{2}{*}{\textbf{Channels/Nodes}} \\ \cline{2-3} \cline{4-5}
                           & Time        & Freq        & Time          & Freq         &                               \\
    \midrule
    CNN1                   & 1           & 7           & 1             & 1            & 128                           \\
    CNN2                   & 7           & 1           & 1             & 1            & 128                           \\
    CNN3                   & 5           & 5           & 2             & 1            & 128                           \\
    CNN4                   & 5           & 5           & 4             & 1            & 128                           \\
    CNN5                   & 5           & 5           & 8             & 1            & 128                           \\
    CNN6                   & 5           & 5           & 16             & 1            & 128                           \\
    CNN7                   & 1           & 1           & 1             & 1            & 8                             \\
    BiLSTM1                & -           & -           & -             & -            & 400                           \\
    BiLSTM2                & -           & -           & -             & -            & 400                           \\
    FC1                    & -           & -           & -             & -            & 601                           \\
    FC2                    & -           & -           & -             & -            & 601  \\                        
    \bottomrule
    \end{tabular}
\end{table}

\begin{figure}[th]
  \centering
  \includegraphics[width=\linewidth]{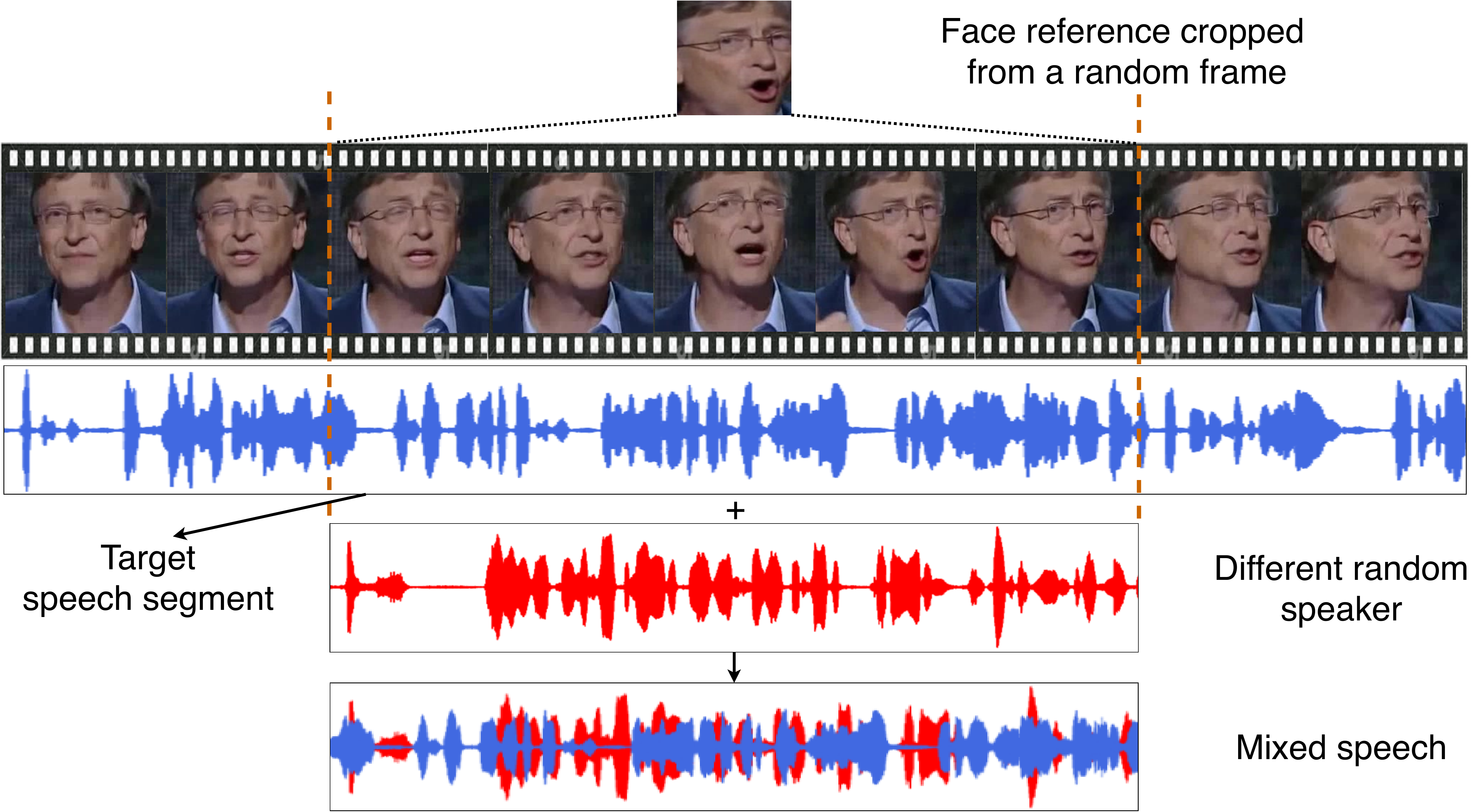}
  \caption{Dataset building.}
  \label{fig:dataset}
\end{figure}

\section{Experimental Setup}
\subsection{Dataset}

We generate the training and test sets based on the Lip Reading Sentences 3 (LRS3) ~\cite{afouras2018lrs3} dataset which consists of thousands of speakers' videos from TED and TEDx. The dataset is transcribed on word-level which will be used in our speech recognition experiments. 

As shown in Figure \ref{fig:dataset}, for the training set, we crop 3s clips from each video where the audio part is treated as the target speech and the visual part is used to get the speaker face from a random frame. To augment the face variants, 10 random faces are extracted from each visual part. The 10 faces are completely out of order and only one face is visible at a time during training. The mixed speech is simulated by directly adding the same length speech from a random different speaker to the target speech. The voice embedding is extracted from a different utterance by the same speaker. Finally, we get 200k samples for around 2k speakers.

For the test set, we use the same process but keep the utterance length in the LRS3 test set and discard the speakers who have only one utterance or there the utterance length is less than 3s. Finally, we get 1171 utterances for 270 speakers. There is no speaker overlap between training and test sets. 

\subsection{Training}
All experiments are conducted on a single NVIDIA Quadro RTX 6000 GPU with 24G memory. We used the Adam optimizer with an initial learning rate of 0.001 and anneal the learning rate with a value of 1.1 after every epoch.

Subsequently, we extract 601-dimension mel-spectrograms with a 25ms window size and a 10ms hop length from mixed speech as model input. Normalization is performed for each mel-frequency bin with the mean and variance.

\subsection{Evaluation metrics}

We evaluate the model performance with two metrics: Source to Distortion Ratio (SDR)~\cite{vincent2006performance} and Word Error Rate (WER). 
SDR\footnote{http://craffel.github.io/mir\_eval/} relates the estimated target signal to the noise terms and was found to negatively correlate with the amount of noise left in the separated audio signal ~\cite{Ephrat_2018}. We also evaluate the signal quality with WER by feeding the separated speech into a ~\textit{Jasper}~\cite{li2019jasper} speech recognition system which is trained on the 960h LibriSpeech dataset and achieves 3.61\% WER on LibriSpeech dev-clean set. The evaluation is performed based on the OpenSeq2Seq\footnote{https://nvidia.github.io/OpenSeq2Seq/html/speech-recognition.html} toolkit  published by NVIDIA.

\begin{figure}[th]
  \centering
  \includegraphics[width=\linewidth]{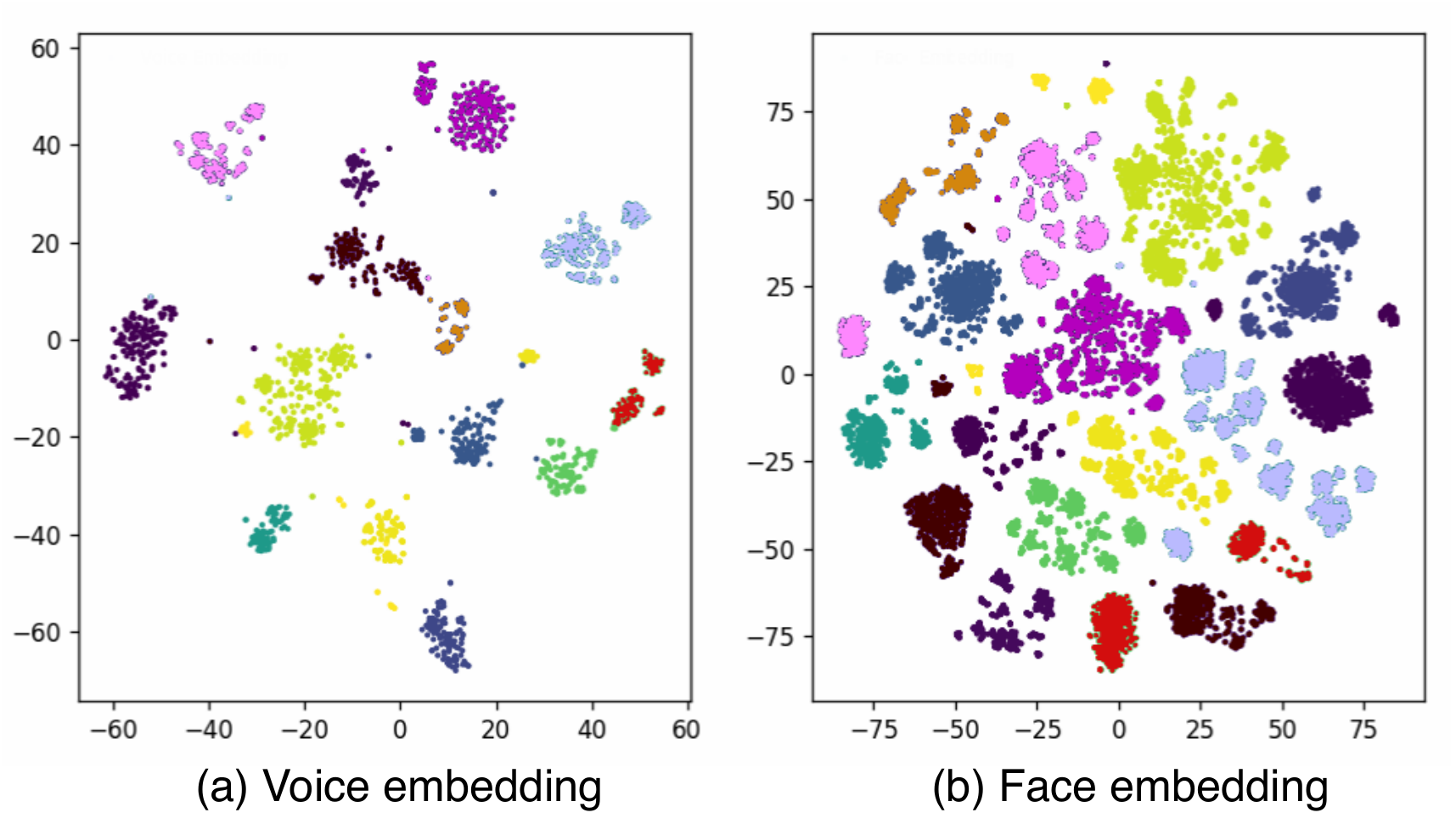}
  \caption{The visualization of (a) voice and (b) face embeddings for 14 randomly chosen speakers in training set with t-SNE.}
  \label{fig:tsne}
\end{figure}

\section{Results and Discussion}
% In this section, we report the model performance on two criterions, source to distortion and word error rate.

\subsection{Results of speech separation}

We visualize the voice and face embeddings from 14 random speakers in the training set. The face embeddings are 10 times more than voices since we randomly crop 10 face images for each mixed speech. As shown in Figure \ref{fig:tsne}, the voice embedding points belonging to the same speakers tend to gather together and significantly far away from other classes. However, the face embedding points from the same speaker are dispersed and close to other classes. We found this is caused by different face angles since all videos are in the wild and the speaker may turn the head from left to right profile while talking.

To investigate the effect of head poses on our experiments, we randomly extract 10 faces for each sample during inference. As shown in Table \ref{tab:sdr}, the performance of using face embeddings fluctuate wildly according to different head poses (Std Dev: 0.32). 

Compared to the result of only using voice embedding (10.32 $\pm$ 0.11 dB), face information achieves competitive performance (9.23 $\pm$ 0.32 dB). The quality of separated speech can be further improved by combining both face and voice references. After checking the output audios, we find that face and voice embeddings are complementary in some cases --- in other words, when two voices sound similar, the corresponding faces may be distinguishable, for example, with different skin colors.

\begin{table}[th]
  \caption{Source to distortion rate results for models using only-voice embedding, only-face embedding and both voice+face embeddings (higher is better).}
  \label{tab:sdr}
  \centering
    \begin{tabular}{cc} \toprule
        {\textbf{Reference}}&  {\textbf{SDR (dB)}}\\ \midrule
        Voice  & 10.32 $\pm$ 0.11 \\ \midrule
        Face  & 9.23 $\pm$ 0.32 \\ \midrule
        Voice+Face  & 10.65 $\pm$ 0.28
        \\ \bottomrule
    \end{tabular}
\end{table}

\subsection{Results of speech recognition}

We test the speech recognition results by Jasper in three settings, clean speech input, mixed speech input and speech separated by our proposed model. The Jasper system achieves 11.8\% WER on the clean inputs, whereas the performance dramatically drops down to 71.2\% WER when using mixed speech input.

We investigate the separated speech inputs for ASR in two conditions. One is the Separated Speech (Clean) in which we test the performance of our proposed model with clean speech input. A robust speech separation system should not only recover desirable output from mixture, but also have good performance on clean speech input. Table 3 lists the similar results for voice (13.46 $\pm$ 0.08\% WER), face (15.31 $\pm$ 0.19\% WER) and
voice+face (13.36 $\pm$ 0.12\% WER) versus clean input (11.83\% WER). The other is the Separated Speech (Mixed) in which the ASR receives the separated speech from mixed signals. We can see, in Table \ref{tab:wer}, the ASR performance can be significantly improved by feeding enhanced speech compared to the 71.22\% WER when directly using noisy speech as input. The speech separation system using voice embedding is superior to the one using face embedding. Combining both voice and face references achieves the lowest WER, which is consistent with the evaluation on SDR.

\begin{table}[th]
  \caption{Word error rate on Jasper speech recognition system.}
  \label{tab:wer}
  \centering
    \begin{tabular}{ccc} \toprule
        {\textbf{Input Speech}}&  {\textbf{Model}}&  {\textbf{WER(\%)}} \\ \cmidrule{1-3}
        Clean Speech  & -  & 11.83  \\ \cmidrule{1-3}
        Mixed Speech  & -  & 71.22  \\ \cmidrule{1-3}
        \multirow{4}{*}{\begin{tabular}[c]{@{}c@{}}Separated Speech\\ (Clean)\end{tabular}} & Voice &13.46 $\pm$ 0.08 \\ \cmidrule(l){2-3}
          & Face & 15.31 $\pm$ 0.19  \\ \cmidrule(l){2-3}
               & Voice+Face &13.36 $\pm$ 0.12 \\ \cmidrule{1-3}
              \multirow{4}{*}{\begin{tabular}[c]{@{}c@{}}Separated Speech\\ (Mixed)\end{tabular}}  & Voice & 25.60 $\pm$ 0.11 \\ \cmidrule(l){2-3}
        & Face & 29.94 $\pm$ 0.25 \\ \cmidrule(l){2-3}
               & Voice+Face & 23.32 $\pm$ 0.12 \\ \bottomrule
    \end{tabular}
\end{table}

\section{Conclusion and Future Work}

In this paper, we propose a novel approach of integrating pre-enrolled face information into the target speech separation task to solve the permutation problem. Different from the conventional audio-visual speech separation methods which heavily rely on the temporal information from the visual sequences, our multimodal system can also be easily adapted to those devices without cameras or to scenarios where no simultaneous visual streams are available. The experimental results on speech separation and speech recognition reveal the effectiveness of face information and the complementarity to voice embeddings.

The face embedding used in our paper is extracted from a model mainly trained on frontal faces (VGGFace2) which is sensitive to the profile views of faces, as indicated in Figure \ref{fig:tsne}.  Future work will focus on adding faces from different angles to the face embedding net training. It is also possible to learn the face embeddings via crossmodal distillation ~\cite{aytar2016soundnet} in which the voice embedding net transfers its knowledge to the face embedding net. This can be applied to scenarios where no voice embedding is available, for instance, a lecture or a colloquium where the clean
speaker voice reference is usually not available, but the speaker face image is accessible on a poster or website.

\section{Acknowledgements}

The authors gratefully acknowledge partial support from the China Scholarship Council (CSC) and from the German Research Foundation DFG under project CML (TRR 169).
\bibliographystyle{IEEEtran}
\bibliography{refs}

\end{document}